# Heat Capacity of Thermally Reduced Graphene Oxide: Compaction and Thermal Annealing Effects

A.I. Krivchikov[1,2], A. Jeżowski[2], M.S. Barabashko[1], A.V. Dolbin[1], N.A. Vinnikov[1], S.V. Cherednichenko[1], Yu. Horbatenko[1], O. Korolyuk[1], O Bezkrovnyi[2], O. Romantsova[1,2], D. Szewczyk[2]*

[1]B. Verkin Institute for Low Temperature Physics and Engineering, NAS of Ukraine, Nauky Ave. 47, 61103 Kharkiv, Ukraine

[2]Institute of Low Temperature and Structure Research PAS, Okólna 2, 50-422 Wrocław, Poland

* Corresponding author. d.szewczyk@intibs.pl

**Abstract**

We present a comprehensive investigation of the low-temperature heat capacity of thermally reduced graphene oxide (trGO) as a function of compaction pressure and annealing temperature. Graphene oxide was synthesized using a modified Hummers method and subsequently thermally reduced at 300 °C, 500 °C, and 700 °C under vacuum to systematically vary the oxygen content and structural ordering. The specific heat data in the 2–300 K range reveal that the thermal response is governed by phonons, including contributions from a Schottky-type anomaly, a defect-related linear term, a Debye term, and a dispersive term with a negative coefficient associated with out-of-plane flexural (ZA) phonons. Increasing compaction pressure alters interlayer coupling and leads to non-monotonic changes in heat capacity, while higher annealing temperatures enhance graphitization, reduce disorder, and modify phonon dispersion. The absence of a boson peak—similar to that observed in carbon nanotubes—supports the dominance of two-dimensional vibrational modes. These findings elucidate the relationship between dimensionality, structural disorder, and processing parameters in shaping the phonon dynamics of trGO, providing guidance for tailoring its thermal behavior in advanced carbon-based functional materials.

# 1. Introduction

Carbon-based materials such as glassy carbon and graphene oxide (GO) have attracted extensive attention for advanced technological applications owing to their outstanding chemical, mechanical, electrical, and thermal properties. These multifunctional characteristics underpin rapid progress in biomedical, pharmaceutical, electronic, and energy-related sectors [1, 2].

Graphene oxide (GO) is a functional derivative of graphene consisting of layered carbon networks decorated with oxygen-containing groups [3]. As the most widely used graphene precursor, GO is also a material with properties distinct from graphene and other carbon allotropes. It is typically synthesized by oxidizing graphite with strong oxidants such as sulfuric acid, potassium permanganate, or hydrogen peroxide [3, 4]. This process introduces hydroxyl and epoxy groups on the basal planes and carbonyl groups at the edges [5]. The resulting functionalization expands the interlayer spacing, enhances hydrophilicity, and promotes dispersibility in water and other solvents [4 – 6]. Depending on the degree of exfoliation, GO can form multilayered or few-layered structures held by π–π interactions [7]. While oxidation disrupts the extended sp² carbon network, producing partially sp³-hybridized domains, it also improves interfacial interactions in composites, albeit at the cost of reduced elastic modulus and electrical conductivity [8, 9].

Thermally reduced graphene oxide (trGO) is obtained by high-temperature treatment of GO, which removes much of the oxygen functionality and partially restores the conjugated sp² lattice [3, 9]. The degree of reduction strongly influences the physical properties of trGO, bridging the gap between easily synthesized GO and high-performance graphene [8, 10]. Compared with GO, trGO exhibits electrical conductivities several times higher due to the reformation of honeycomb networks, breaking of oxygen–carbon bonds, and recovery of π-electron delocalization [9, 10].

At low temperatures, the heat capacity of carbon-based solids is primarily determined by phonon excitations, with the temperature dependence strongly influenced by dimensionality, crystallinity, and disorder [11]. In crystalline graphite, the Debye model predicts a cubic ($T^3$) behavior characteristic of a well-ordered phonon density of states [12]. In contrast, reduced dimensionality or structural disorder—as in graphene nanoplatelets, trGO, and carbon nanotubes (CNTs)—leads to deviations from Debye scaling [13, 14]. For instance, isolated CNTs are predicted to exhibit a linear heat capacity at very low $T$ due to one-dimensional acoustic modes, whereas few-layer graphene may follow a quadratic ($T^2$) law [12, 15]. Other graphitic systems such as expanded graphite (EG) and mesophase pitch–based carbon fibers (MPCFs) display even more complex behavior due to anisotropy, finite crystallite size, and grain-boundary imperfections, which generate additional low-frequency vibrational states [16 – 20].

Thermally reduced graphene oxide is of particular interest because, unlike pristine graphene, it retains substantial structural disorder from residual oxygen groups, defects, wrinkles, and porosity. These features modify its phonon spectrum and introduce excess contributions to the heat capacity, often exceeding Debye predictions [21, 22]. In some cases, a linear-in-T term appears, resembling the behavior of glasses and other disordered solids [23].

Experimental calorimetry, complemented by Raman spectroscopy, X-ray diffraction, and electron microscopy, provides critical insights into how defect density, crystallite size, interlayer coupling, and morphological features influence the phonon density of states.

Studying the thermal properties of trGO is therefore essential. Its disordered nanostructure fundamentally alters vibrational dynamics compared to pristine graphene, impacting both thermal conductivity and low-temperature heat capacity. Understanding these effects is vital for tailoring trGO in thermal management applications, such as heat dissipation in electronic devices and energy storage systems. Furthermore, analysis of its low-temperature specific heat provides valuable insights into low-energy excitations, including two-level systems, that are hallmarks of disordered solids. These investigations thus bridge the gap between theory and experiment in nanoscale thermal transport.

## 2. Materials and experimental methods

Graphene oxide (GO) was synthesized from high-purity graphite powder using the modified Hummers method [24, 25]. In this process, oxygen-containing functional groups (hydroxyl, epoxy, and carboxyl) were covalently grafted onto the graphite lattice, disrupting the extended $sp^2$ network and increasing the interlayer spacing. This transformation partially altered the carbon hybridization from planar $sp^2$ toward a more tetrahedral $sp^3$ configuration [26].

Thermal exfoliation was subsequently employed to delaminate the oxidized graphite into graphene layers. Under vacuum conditions ($\sim 10^{-3}$ mmHg), the samples were slowly heated (5 – 7°C/min) up to ~300°C and at this the temperature the sample was kept for 15 min, after which the cell was cooled to room temperature for one and a half hour under vacuum [23]. The release of intercalated water and labile oxygen groups generated internal pressure, driving the exfoliation of the oxidized graphite. The resulting material, designated trGO_300 (samples S1÷S6), was characterized by energy-dispersive X-ray spectroscopy (EDS), which revealed ~83.6 at.% carbon, ~14 at.% oxygen, and <2.4 at.% impurities [23, 27 – 29].

Further thermal treatments at higher temperatures yielded more reduced forms of graphene oxide. Heating trGO_300 under vacuum ($\sim 10^{-3}$ mmHg) at 500°C and 700°C for 30 minutes produced trGO_500_S6 and trGO_700_S6, respectively. Successive reduction decreased oxygen content and promoted greater graphitic ordering, though structural defects and disorder persisted [30].

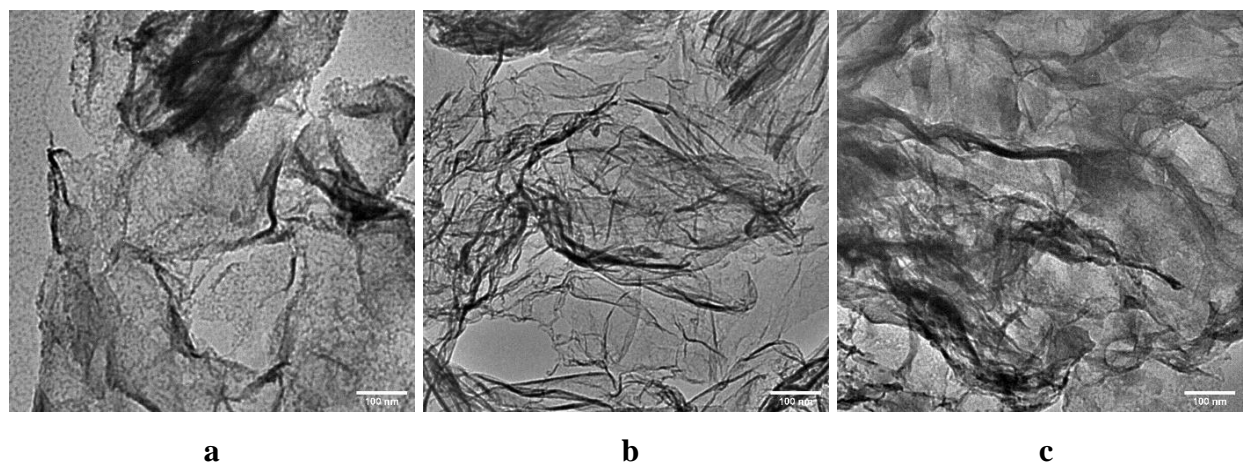

|  a  |  b  |  c  |

**Fig. 1.** TEM images of powder sample S6 of the trGO_300 (a), trGO_500 (b) and trGO_700 (c), which were obtained by thermo-vacuum exfoliation at different temperatures: 300°C, 500°C, and 700°C, respectively.

**Figure 1** presents TEM images of trGO samples obtained at 300°C (a), 500°C (b), and 700°C (c). These images document the pronounced morphological evolution of trGO—from loosely packed, single-layer flakes at 300°C to increasingly compact, wrinkled, and aggregated structures at higher reduction temperatures.

- **trGO_300** (Fig. 1a) retains a high density of oxygen groups and possibly residual intercalated water. Localized dark contrasts (1–10 nm) on the surfaces correspond to covalently attached oxygen species (–OH, –COOH), which disrupt sp² hybridization and alter local electronic density.
- **trGO_500** (Fig. 1b), shows pronounced bending and folding of the graphene sheets. Approximately 60–70% of oxygen groups and all intercalated water are removed at this stage [30]. The detachment of oxygen atoms from the carbon plane produces structural defects (e.g., vacancies, pentagon–heptagon reconstructions) and induces partial sp²→sp³ rehybridization, which contributes to the observed wrinkling.
- **trGO_700** (Fig. 1c), exhibits up to ~90% removal of oxygen groups [30]. The carbon sheets stack more closely, reducing the interlayer distance and giving rise to dark, strongly absorbing regions in the TEM images, consistent with locally "graphitized" domains.

These TEM observations highlight how progressive thermal reduction modifies both the chemistry and nanomorphology of GO, with direct implications for its vibrational spectrum and low-temperature heat capacity.

For calorimetric studies, bulk samples were prepared by pressing trGO powders into pellets 10 mm in diameter and 2–3 mm thick. Most samples were compacted under 1.25 GPa, while set S5 was prepared at three different pressures (1.25, 1.0, and 0.75 GPa) to evaluate compaction effects. The commercial setup, Physical Properties Measurement System (PPMS, Quantum Design Inc.), was used to determine the heat capacity by the thermal relaxation method in the temperature range 1.8–275 K. The random experimental error did not exceed 2–3%.

## 3. Results and discussion

**Figure 2** shows the temperature dependence of the heat capacity $C_p(T)$ for different series of trGO_300, trGO_500, and trGO_700, and earlier data of trGO_300 by Sumarokov et al. [14]. A good agreement is observed above 5 K, where phonons dominate. However, below 5 K, the literature data [14] of trGO_300 lie higher than our present data. At 2 K the heat capacity values $C_p(T)$ is approximately twice that of trGO_300 (literature to S2 sample comparison). These discrepancies highlight the sensitivity of the low-temperature heat capacity to synthesis routes, oxidation conditions, and structural morphology.

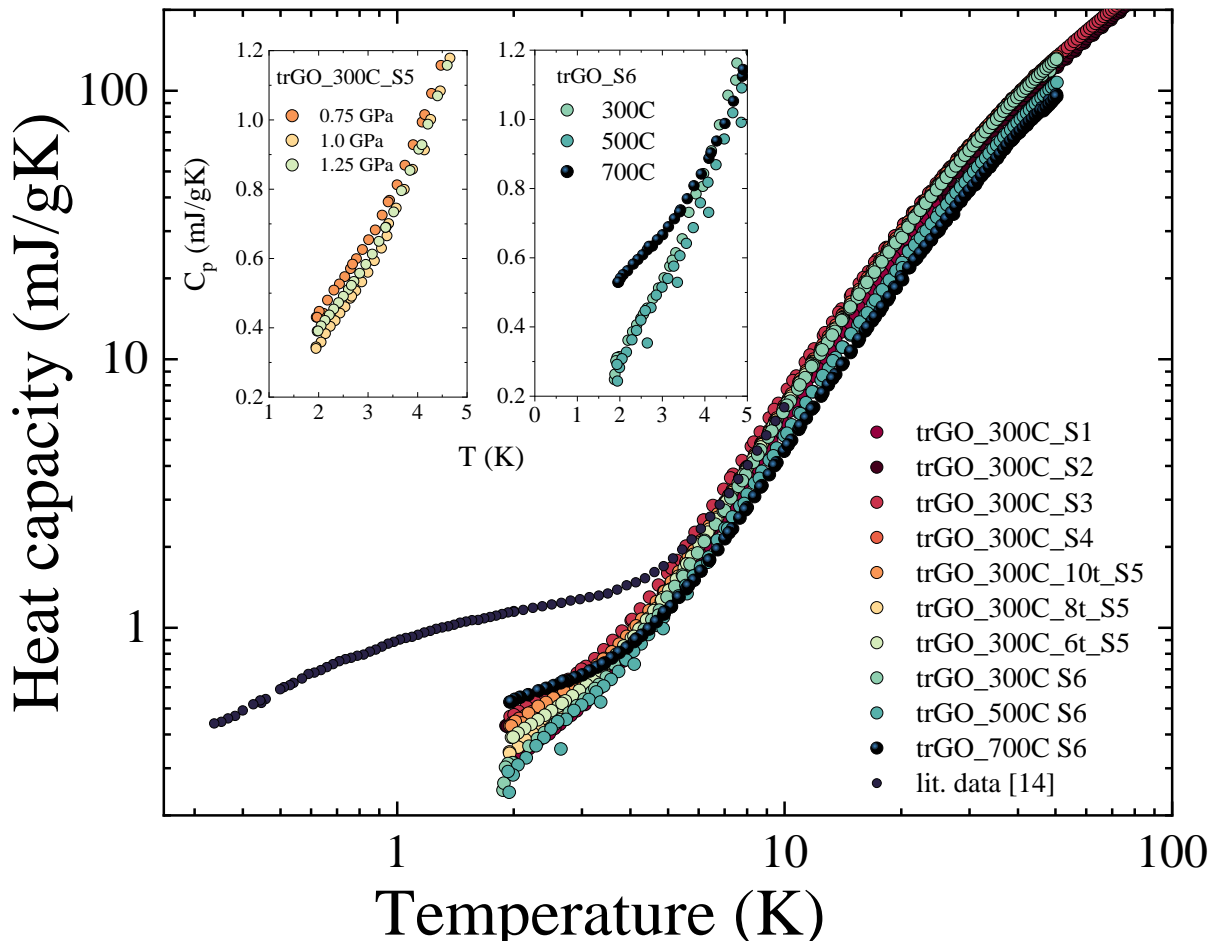

**Fig. 2.** Temperature dependences of heat capacity for different trGO-samples. Insets: close-up on the compacted (S5) and annealed (S6) samples behavior at the lowest temperatures.

The specific heat of carbon-based materials is intrinsically linked to their density, with lower-density structures exhibiting increased heat capacity across both ordered (crystalline) and disordered (amorphous) phases. [31, 32]. The left inset in Fig. 2 shows that below 4 K, the set of three trGO_300 S5 samples obtained with different pressures demonstrates exhibit different heat-capacity behavior. This observation reflects the 2D structure and high elastic modulus of trGO [8, 9], which produce an anisotropic response during compression. Under moderate pressure, trGO sheets appear to stack more tightly, reducing interlayer spacing. Thus, compacting trGO results in a transition toward a more graphitized structure [33], accompanied by an increase in the mean size of ordered domains.

The remarkable thermal characteristics of carbon-based materials continue to be a subject of considerable scientific interest [34, 35]. Both theoretical analyses and experimental investigations of low-dimensional systems consistently demonstrate that phonons are the dominant contributors to heat capacity at low temperatures [11, 14, 15, 36]. The temperature dependence of $C_p$ represented as $C_p/T$ versus $T^2$ is illustrated in **Figure 3a**. Four principal components contributing to the total heat capacity have been identified:
- a Schottky-like contribution ($C_S$),
- a linear term ($A_1T$),
- a Debye phonon contribution ($A_3T^3$),
- a negative "flexible" term ($A_5T^5$).

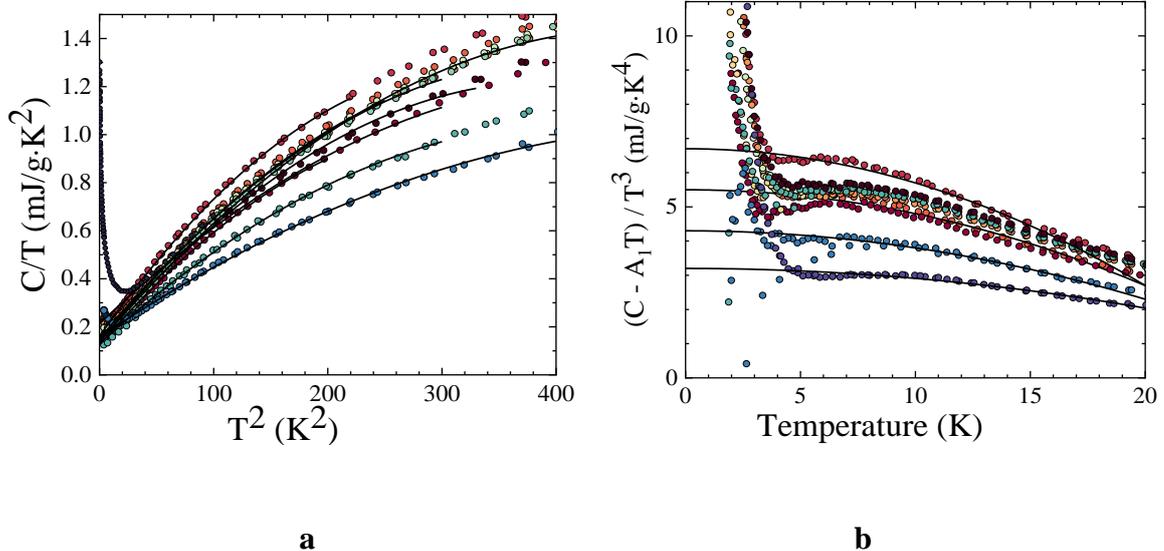

a  b

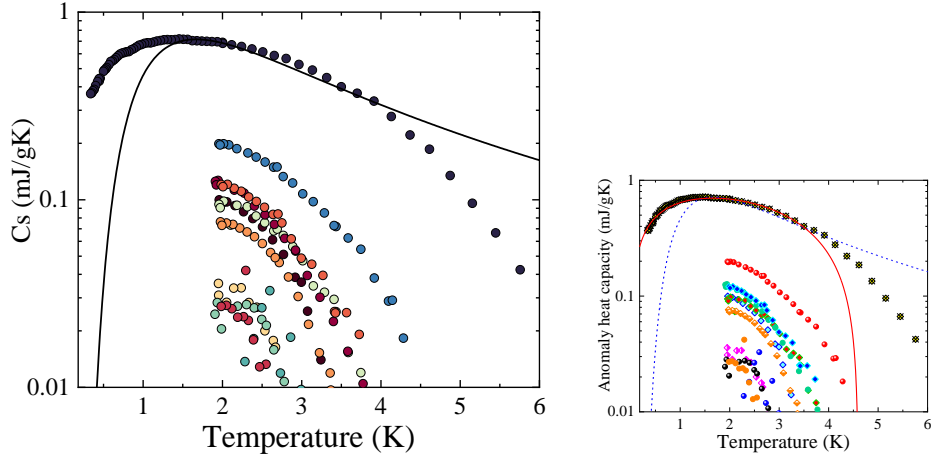

**Fig. 3.** Temperature dependences of heat capacity for different trGO-samples as $C/T$ vs $T^2$ (a) and as $(C-A_1T)/T^3$ vs $T$ (b) and Schottky-like contribution $C_S$ vs $T$ (c). Symbols are the same as in Fig.2. Extrapolation of the $C_{ph}(T)$ based on the eq.1 presented by the solid lines Fig. 3(a, b). The calculated Schottky heat capacity for the two-level system with the energy gap of 4 K presented by the solid line on Fig. 3(c).

At temperatures below 10 K, the experimental data are well described by the following expression:

$$C(T) = C_S(T) + A_1T + C_{ph}(T), \quad (1)$$

where $C_{ph}(T) = A_3T^3 + A_5T^5$ are corresponding to the low-temperature series expansion typically observed in carbon-based materials [11] with fitting parameters $A_1$, $A_3$, and $A_5$. The particular parameters obtained from fitting the experimental data with equation 1 are summarized in **Table 1**. The negative $A_5$ term arises from the contribution of the out-of-plane flexural (ZA) phonon mode, whose behavior in the long-wavelength limit is governed by the following dispersion relation:

$$\omega^2(q) = s^2q^2 + k^2q^4, \quad (2)$$

with $s$ denoting the sound velocity and $k$ the flexural rigidity parameter [37, 38]. Wrinkling and structural defects intensify phonon scattering, leading to deviations from the conventional Debye law ($C \sim T^3$) as elevated temperatures. In monolayer and few-layer trGO, heat capacity exhibits a modified two-dimensional phonon behavior, diverging notably from that observed in bulk graphite [12, 15].

The linear term $A_1T$ in eq. (1) may may originate from several mechanisms, including two-dimensional phonon modes in reduced graphene oxide [39], two-level systems (TLS) associated with structural disorder [40], or intrinsic defects within carbon nanostructures [41, 42]. Ripples and topological irregularities give rise to localized vibrational states, while short-wavelength acoustic and optical phonons contribute to the linear temperature dependence of the heat capacity [42].

The plot of $C(T) - A_1T$ giving the lattice and Schottky-like contributions to the total heat capacity, is presented in **Figure 3b** as $C_{ph}/T^3$ versus T. The solid lines correspond to fitted phonon heat capacity curves $C_{ph}/T^3 = A_3 + A_5T^2$, and show at T>4 K strong agreement with the experimental data. For thermally reduced graphene oxide, the extrapolated $C_{ph}/T^3$ remains nearly constant below 2 K, whereas it decreases with increasing temperature—closely mirroring the behavior observed in graphite and carbon nanotubes [11]. This trend reflects the layered architecture of trGO, where pronounced bonding anisotropy influences phonon dynamics. Notably, the boson peak typically associated with disordered solids is absent, a feature attributed to the negative $A_5$ term. Comparable behavior has been reported for single-walled and multi-walled carbon nanotubes (SWCNTs and MWCNTs), where the negative high-order contributions arise from the out-of-plane ZA (flexural) phonon modes, described by the dispersion relation given in Equation (2) [11].

The heat capacity additional contribution Cs corresponds to the Schottky-type anomaly similar to that reported in [14]. As temperature decreases, Cs increases and tends toward saturation. The values of Cs at T≈2K are presented in **Table 1**. Notably, Cs depends strongly on the degree of compaction of the trGO material (see **Figure 3c**). This compaction effect arises from interlayer contact phenomena. Therefore, the observed anomaly can be regarded as a characteristic fingerprint of trGO, directly linked to layer–layer interactions and compaction effects.

Annealing of the trGO (sample S6) leads to irreversible structural transformations (graphitization), clearly reflected in the calorimetric data (right inset in Fig. 2), e.g. by changes in fitting parameters - **Table 1** or **Figure 4a** – showing the phonon contribution dependence on annealing temperature.

**Tab. 1.** Parameters of the fitting of the heat capacity data for the samples of trGO: m – sample mass, $A_1$, $A_3$, $A_5$ fitting parameters according to equation (1), $C_s$ – value of the Schottky-like contribution at 2K.

| Samples | m [mg] | $A_1$ [mJ/g·K$^2$] | $A_3$ [μJ/g·K$^4$] | $A_5$ [μJ/g·K$^6$] | $C_S$ [mJ/g·K] |
|---|---|---|---|---|---|
| trGO_300_S1 | 1.32 | 0.12 | 5.5 | -0.008 | 0.027 |
| trGO_300_S2 | 3.82 | 0.14 | 5.5 | -0.007 | 0.124 |
| trGO_300_S3 | 2.63 | 0.16 | 6.7 | -0.01 | 0.10 |
| trGO_300_S4 | 2.56 | 0.14 | 5.9 | -0.0075 | 0.12 |
| trGO_300_1.25GPa_S5 | 3.11 | 0.15 | 5.7 | -0.007 | 0.10 |
| trGO_300C_1GPa_S5 | 3.64 | 0.14 | 5.9 | -0.0075 | 0.036 |
| trGO_300C_0.75GPa_S5 | 3.51 | 0.14 | 5.9 | -0.0075 | 0.074 |
| trGO_300_S6 | 5.60 | 0.12 | 5.9 | -0.0075 | 0.027 |
| trGO_500_S6 | 8.85 | 0.13 | 4.3 | -0.005 | 0.032 ~~±0.01~~ |
| trGO_700_S6 | 1.45 | 0.16 | 3.2 | -0.0029 | 0.20 |

The Schottky-type anomaly observed in trGO is therefore not only a characteristic feature but also a sensitive probe of structural evolution. **Figure 4a** and **Table 1** demonstrate that both the Debye phonon ($A_3T^3$) and the negative flexible ($A_5T^5$) contributions are strongly affected by annealing. As the annealing temperature increases, $A_3$ decreases, and both terms show a dramatic reduction in heat capacity. The linear term $A_1$ changes from 0.12 up to 0.16 mJ/g·K$^2$. These changes reflect the transformation of trGO toward a more ordered, graphite-like structure.

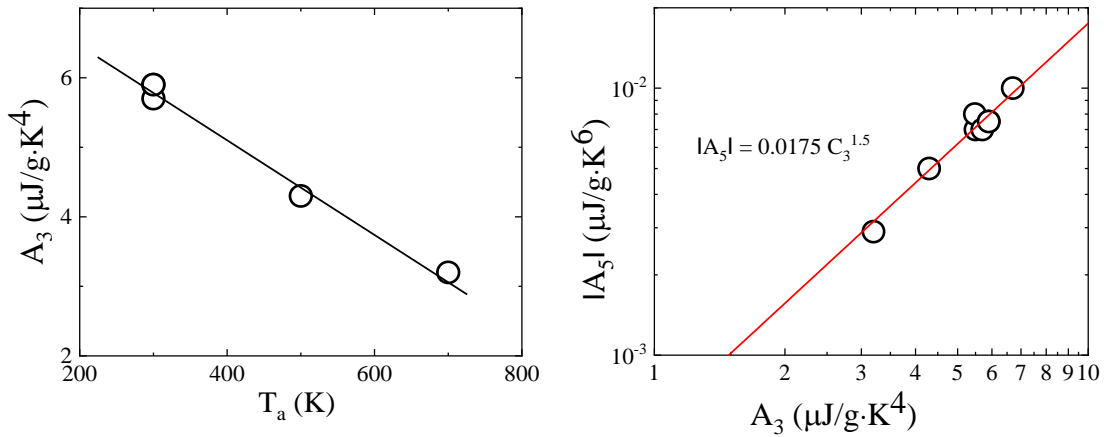

**Fig. 4.** The relationship between heat capacity parameters: $A_3$ vs $T_a$, where $T_a$ – temperature of the annealing of the trGO and $A_5$ vs $A_3$. Solid lines demonstrate linear dependence.

Finally, **Figure 4b** reveals a power-law correlation between the Debye and flexible contributions, indicating their coupled evolution with annealing and wrinkle formation. Similar behavior has been reported for carbon nanotubes [11], where the weakening of flexural phonons accompanies an increase in ordered stacking and graphitization.

**Conclusions**

Low-temperature calorimetric measurements of thermally reduced graphene oxide (trGO) with varying compaction pressures and annealing temperatures reveal that its heat capacity is governed by a combination of phononic and defect-related contributions. The data are well described by a four-term expression incorporating a Schottky-type anomaly, a linear term, a Debye term, and a negative flexural term associated with out-of-plane ZA phonons.

Compaction pressure produces non-monotonic changes in heat capacity below 4 K, reflecting the anisotropic mechanical response of the layered structure and variations in interlayer coupling. The Schottky-type anomaly emerges as a sensitive fingerprint of layer–layer interactions, also slightly dependent on compaction.

Thermal annealing induces irreversible structural transformations toward a more graphitized state, reducing disorder and modifying phonon dynamics. This is manifested in systematic changes to the Debye and flexible phonon contributions, as well as in the suppression of the boson peak typically observed in disordered solids.

The observed correlations between structural order, dimensionality, and vibrational behavior establish clear processing–structure–property relationships for trGO. These insights provide a framework for tailoring the thermal properties of graphene-based nanomaterials for applications in electronics, thermal management, and energy systems.

**Funding:**

This work was partly supported by the National Research Foundation of Ukraine (Grant 2023.03/0012) and National Science Centre Poland (Grant 2022/45/B/ST3/02326).

*Reference:*

[1] de Souza Vieira, L. (2022). A review on the use of glassy carbon in advanced technological applications. *Carbon*, *186*, 282-302, https://doi.org/10.1016/j.carbon.2021.10.022.

[2] Lu, Y., Li, H., Rong, J., Yu, X., Sui, Y., & Liu, Z. (2025). An effective strategy for enhancing 2D graphene-like thermoelectric performance: Hydrogenation distortion. *Diamond and Related Materials*, *154*, 112209, https://doi.org/10.1016/j.diamond.2025.112209.

[3] Wu, J., Lin, H., Moss, D. J., Loh, K. P., & Jia, B. (2023). Graphene oxide for photonics, electronics and optoelectronics. Nature Reviews Chemistry, 7(3), 162-183, https://doi.org/10.1038/s41570-022-00458-7 .

[4] Razaq, A., Bibi, F., Zheng, X., Papadakis, R., Jafri, S. H. M., & Li, H. (2022). Review on graphene-, graphene oxide-, reduced graphene oxide-based flexible composites: From fabrication to applications. *Materials*, 15(3), 1012, https://doi.org/10.3390/ma15031012 .

[5] Aliyev, E., Filiz, V., Khan, M. M., Lee, Y. J., Abetz, C., & Abetz, V. (2019). Structural characterization of graphene oxide: Surface functional groups and fractionated oxidative debris. *Nanomaterials*, 9(8), 1180, https://doi.org/10.3390/nano9081180.

[6] Yu, H., He, Y., Xiao, G., Fan, Y., Ma, J., Gao, Y., ... & Mei, X. (2020). The roles of oxygen-containing functional groups in modulating water purification performance of graphene oxide-

based membrane. *Chemical Engineering Journal*, 389, 124375, https://doi.org/10.1016/j.cej.2020.124375.

[7] Acik, M., & Chabal, Y. J. (2013). A review on thermal exfoliation of graphene oxide. *Journal of Materials Science Research*, 2(1), 101, doi:10.5539/jmsr.v2n1p101 .

[8] Singh, A., & Kumar, D. (2019). Effect of functionalization on the elastic behavior of graphene nanoplatelet-PE nanocomposites with interface consideration using a multiscale approach. *Mechanics of Materials*, 132, 18-30, https://doi.org/10.1016/j.mechmat.2019.02.008.

[9] Trikkaliotis, D. G., Christoforidis, A. K., Mitropoulos, A. C., & Kyzas, G. Z. (2021). Graphene oxide synthesis, properties and characterization techniques: a comprehensive review. *ChemEngineering*, 5(3), 64, https://doi.org/10.3390/chemengineering5030064.

[10] Zhang, P., Li, Z., Zhang, S., & Shao, G. (2018). Recent advances in effective reduction of graphene oxide for highly improved performance toward electrochemical energy storage. *Energy & Environmental Materials*, 1(1), 5-12, https://doi.org/10.1002/eem2.12001.

[11] Barabashko, M. S., Krivchikov, A. I., Jeżowski, A., Bezkrovnyi, O., Bagatskii, M. I., Sumarokov, V. V., ... & Szewczyk, D. (2025). Experimental Evidence of Flexural Phonons in Low-Temperature Heat Capacity of Carbon Nanotubes. *Carbon Trends*, 100479, https://doi.org/10.1016/j.cartre.2025.100479.

[12] Viana, R., Godfrin, H., Lerner, E., & Rapp, R. (1994). Unusual specific-heat contribution in exfoliated graphite. *Physical Review B*, 50(7), 4875–4877, DOI: https://doi.org/10.1103/PhysRevB.50.4875 .

[13] Barabashko, M., & Nakazawa, Y. (2021). Experimental heat capacity of low-dimensional systems: Carbon nanotubes and 1D atomic and molecular chains of adsorbates. *Netsu Sokutei*, 48(4), 164-170, https://doi.org/10.11311/jscta.48.4_164.

[14] Sumarokov, V. V., Dolbin, A. V., Jeżowski, A., Szewczyk, D., Gnida, D., Vinnikov, N. A., & Bagatskii, M. I. (2024). Short notes: Measurements on the heat capacity of thermal reduced graphene oxide down to 0.3 K. *Low Temperature Physics*, 50(2), 185-187, https://doi.org/10.1063/10.0024332.

[15] Gospodarev, I. A., Sirenko, V. A., Syrkin, E. S., Feodosyev, S. B., & Minakova, K. A. (2020). Low dimensional features of graphene nanostructure stability and vibrational characteristics (Review). *Low Temperature Physics*, 46(3), 232–257, https://doi.org/10.1063/10.0000706.

[16] Inagaki, M., Kaburagi, Y., & Hishiyama, Y. (2014). Thermal Management Material: Graphite. Advanced Engineering Materials, 16(5), 494–506, https://doi.org/10.1002/adem.201300418

[17] Emmerich, F. G. (2014). Young's modulus, thermal conductivity, electrical resistivity and coefficient of thermal expansion of mesophase pitch-based carbon fibers. *Carbon*, 79, 274–293, https://doi.org/10.1016/j.carbon.2014.07.068Get rights and content

[18] Chen, T., Huang, Y., Wei, L., Xu, T., & Xie, Y. (2023). Thermal and electrical transport in partly-reduced graphene oxide films: The effect of low temperature and structure domain size. *Carbon*, 203, 130-140, https://doi.org/10.1016/j.carbon.2022.11.051.

[19] Bodzenta, J., Mazur, J., & Kaźmierczak-Bałata, A. (2011). Thermal properties of compressed expanded graphite: photothermal measurements. *Applied Physics B*, 105, 623-630, https://doi.org/10.1007/s00340-011-4510-7 .

[20] Zorzi, J. E., & Perottoni, C. A. (2021). Thermal expansion of graphite revisited. *Computational Materials Science*, 199, 110719, https://doi.org/10.1016/j.commatsci.2021.110719.

[21] Ding, G., Duan, J., Cai, S. L., Dai, L. H., & Jiang, M. Q. (2025). Universal correlation between boson peak and quasi-localized modes in solids. *Journal of Non-Crystalline Solids*, 666, 123668, https://doi.org/10.1016/j.jnoncrysol.2025.123668.

[22] Szewczyk, D., Krivchikov, A. I., Barabashko, M. S., Dolbin, A. V., Vinnikov, N. A., Basnukaeva, R., ... & Jeżowski, A. (2023). Universal behavior of low-temperature heat capacity of acrylonitrile-butadiene-styrene thermoplastic polymer and its composite with graphene oxide. *Low Temperature Physics*, 49(5), 593-593, https://doi.org/10.1063/10.0017821.

[23] Sumarokov, V. V., Jeżowski, A., Szewczyk, D., Dolbin, A. V., Vinnikov, N. A., & Bagatskii, M. I. (2020). The low-temperature specific heat of thermal reduced graphene oxide. *Low Temperature Physics*, 46(3), 301-305, https://doi.org/10.1063/10.0000703.

[24] Zhou, F., Fathizadeh, M., & Yu, M. (2018). Single-to few-layered, graphene-based separation membranes. *Annual Review of Chemical and Biomolecular Engineering*, 9(1), 17-39, https://doi.org/10.1146/annurev-chembioeng-060817-084046.

[25] H.V. Rusakova, L.S. Fomenko, S.V. Lubenets, A.V. Dolbin, M.V. Khlistyuck, A.V. Blyznyuk, (2020), Synthesis and micromechanical properties of graphene oxide-based polymer nanocomposites, *Low Temperature Physics* 46 (3), 276-284, https://doi.org/10.1063/10.0000699.

[26] Zhang, C., Lv, W., Xie, X., Tang, D., Liu, C., & Yang, Q. H. (2013). Towards low temperature thermal exfoliation of graphite oxide for graphene production. *Carbon*, 62, 11-24, https://doi.org/10.1016/j.carbon.2013.05.033.

[27] A. V. Dolbin, N. A. Vinnikov, V. B. Esel'son et al., (2020) The impact of treating graphene oxide with a pulsed high-frequency discharge on the low-temperature sorption of hydrogen, *Low Temperature Physics* 46, 293, https://doi.org/10.1063/10.0000701

[28] Barabashko, M. S., Drozd, M., Dolbin, A. V., Basnukaeva, R. M., & Vinnikov, N. A. (2024). Kinetics of the thermal decomposition of thermally reduced graphene oxide treated with a pulsed high-frequency discharge in the hydrogen atmosphere. *Low Temperature Physics*, 50(5): 368-371, https://doi.org/10.1063/10.0025619.

[29] Dolbin, A.V., Vinnikov, N. A., Esel'son, V. B., Gavrilko, V. G., Basnukaeva, R. M., Khlistuck, M. V., Maser, W. K., Benito; A. M. (2019) The effect of graphene oxide reduction temperature on the kinetics of low-temperature sorption of hydrogen. *Low Temperature Physics* 45(4): 422–426, https://doi.org/10.1063/1.5093523.

[30] Dolbin, A.V., Khlistuck, M.V., Esel'son, V. B., Gavrilko, V. G., Vinnikov, N. A., Basnukaeva, R. M., Maluenda, I., Maser, W. K., Benito; A. M. (2015) The effect of the thermal reduction temperature on the structure and sorption capacity of reduced graphene oxide materials, *Applied Surface Science* 361,213-220, https://doi.org/10.1016/j.apsusc.2015.11.167.

[31]. Barabashko, M. S., & Krivchikov, A. I. (2025). About the hump in the low-temperature isochoric heat capacity of Ne cryocrystals. *Low Temperature Physics*, 51(2), 157-161, https://doi.org/10.1063/10.0034896.

[32] Chumakov, A. I., & Monaco, G. (2015). Understanding the atomic dynamics and thermodynamics of glasses: Status and outlook. *Journal of Non-Crystalline Solids*, 407, 126–132, https://doi.org/10.1016/j.jnoncrysol.2014.09.031.

[33] Chen, X., Deng, X., Kim, N. Y., Wang, Y., Huang, Y., Peng, L., … Ruoff, R. S. (2018). Graphitization of graphene oxide films under pressure. *Carbon*, 132, 294–303, https://doi.org/10.1016/j.carbon.2018.02.049.

[34] Moon, J., & Tian, Z. (2025). Crystal-like thermal transport in amorphous carbon. *npj Computational Materials*, *11*(1), 137, https://doi.org/10.1038/s41524-025-01625-2 .

[35] Un, H. I., Iwanowski, K., Ferrer Orri, J., Jacobs, I. E., Fukui, N., ... & Nishihara, H. Sirringhaus, H. (2025). Defect-tolerant electron and defect-sensitive phonon transport in quasi-2D conjugated coordination polymers. *Nature Communications* 16, 6628, https://doi.org/10.1038/s41467-025-61920-w .

[36] Davydov, V. N. (2020). Role of the Lifshitz topological transitions in the thermodynamic properties of graphene. *RSC advances*, 10(46), 27387-27400, 10.1039/D0RA04601A .

[37] Gospodarev, I. A., Grishaev, V. I., Manzhelii, E. V., Syrkin, E. S., Feodosyev, S. B., & Minakova, K. A. (2017). Phonon heat capacity of graphene nanofilms and nanotubes. *Low Temperature Physics*, 43(2), 264-273, https://doi.org/10.1063/1.4978291.

[38] Popov, V. N. (2002). Low-temperature specific heat of nanotube systems. *Physical Review B*, 66(15), DOI: https://doi.org/10.1103/PhysRevB.66.153408 .

[39] A. Taheri, et al., Importance of quadratic dispersion in acoustic flexural phonons for thermal transport of two-dimensional materials, *Phys. Rev. B* 103 (23) (2021) 235426, DOI: https://doi.org/10.1103/PhysRevB.103.235426 .

[40] Szewczyk, D., Moratalla Martín, M. E., Chajewski, G., Gebbia, J. F., Jezowski, A., Krivchikov, A., ... & Ramos Ruiz, M. A. (2024). Specific heat at low temperatures in quasiplanar

molecular crystals: origin of glassy anomalies in minimally disordered crystals. *Physical Review. B*, 110, 174204, DOI: https://doi.org/10.1103/PhysRevB.110.174204

[41] Barabashko, M. S., Rezvanova, A. E., & Ponomarev, A. N. (2017). Low temperature heat capacity and sound velocity in fullerite C60 orientational glasses. *Fullerenes, Nanotubes and Carbon Nanostructures*, 25(11), 661-666, https://doi.org/10.1080/1536383X.2017.1391225

[42] Cano, A., & Levanyuk, A. P. (2004). Low-temperature specific heat of real crystals: Possibility of leading contribution of optical vibrations and short-wavelength acoustical vibrations. *Physical Review B—Condensed Matter and Materials Physics*, 70(21), 212301, https://doi.org/10.1103/PhysRevB.70.212301